\documentstyle[12pt,aps,prd,preprint]{revtex}
\begin{document}
\title{Cosmic Censorship, Area Theorem, and Self-Energy of Particles}
\author{Shahar Hod}
\address{Department of Condensed Matter Physics, Weizmann Institute, 
Rehovot 76100, Israel}
\date{\today}
\maketitle

\begin{abstract} 

The (zeroth-order) energy of a particle in the background of a black hole 
is given by Carter's integrals. However, exact calculations of a particle's 
{\it self-energy} (first-order corrections) are still beyond our present reach in many situations. 
In this paper we use Hawking's area theorem in order 
to derive bounds on the self-energy of a particle in the vicinity of a black hole. Furthermore, we 
show that self-energy corrections {\it must} be taken into account in order to guarantee the validity of 
Penrose cosmic censorship conjecture.

\end{abstract}

\section{Introduction}

Spacetime singularities that arise in gravitational collapse are
always hidden inside of black holes. 
This is the essence of the (weak) cosmic censorship conjecture, 
put forward by Penrose thirty years ago \cite{Pen}. 
The conjecture, which is widely believed to be true, has become one of the corner stones of 
general relativity. Moreover, it is being envisaged as a
basic principle of nature.  However, despite the flurry of activity
over the years, the validity of this conjecture is still an open
question (see e.g., \cite{Wald1,Sin,Wald2,His,BekRos,Hub,QuWa,Hod1,HodPir} and references therein). 

The destruction of a black-hole event horizon is ruled out by this
principle because it would expose the inner singularities to distant
observers. Moreover, the horizon area of a black hole, $A$, 
is associated with an entropy $S_{BH}=A/4\hbar$ (we use gravitational units in which $G=c=1$). 
Thus, without any obvious physical mechanism to compensate for the
loss of the black-hole enormous entropy, 
the destruction of the black-hole event horizon would 
violate the generalized second law (GSL) of thermodynamics
\cite{Beken1}. For these two reasons, any process which seems, at
first sight, to remove the black-hole 
horizon is expected to be unphysical. For the advocates of the cosmic
censorship principle the task remains to find out how such candidate
processes eventually fail to remove the horizon.

As is well-known, the Kerr-Newman metric with $M^2 -Q^2 -a^2 < 0$ (where $M, Q,$ and $a$ are the
mass, charge and specific angular momentum, respectively) does not contain an event
horizon, and it therefore describes a naked singularity. Thus, one may try 
to ``over-spin'' (or ``over-charge'') a black hole 
by dropping into it a rotating (or a charged) particle. Such gedanken experiments allow one to 
test the consistency of the conjecture. 
It turns out that the {\it test} particle approximation actually allows 
a black hole to ``jump over'' extremality in 
this type of gedanken experiments. We show that in order to guarantee the integrity of the 
black-hole event horizon one {\it must} take into account the {\it self-energy} of the particle 
(first-order interactions between the black hole and the object). 

Furthermore, a well-established theorem in the physics of black holes is 
Hawking's area theorem \cite{Haw}, according to which the black-hole surface
area should increase (or remains unchanged) in such gedanken experiments. We show that it 
is possible to use the area theorem in order to derive bounds on the self-energy of particles in 
the a black hole spacetime. 
It should be further noted that 
in recent years there is a growing interest in the calculation of the self-interaction of a particle in the 
spacetime of a black hole (see e.g., \cite{NaMiSa} and references therein). 
The flurry of activity in this area of research is motivated by the prospects of 
detection of gravitational waves in the future by gravitational wave detectors such as LISA \cite{Lisa}.

\section{Self-energy of a particle with angular momentum}

We consider a particle 
which is lowered towards an extremal Kerr black hole. 
To {\it zeroth} order in particle-hole interaction 
the energy (energy-at-infinity), ${\cal E}^{(0)}$, of the object in the black-hole             
spacetime is given by Carter's \cite{Carter}
integrals (constants of motions). As shown by Christodoulou
\cite{Chris} (see also \cite{ChrisRuff}), 
${\cal E}^{(0)}(r=r_+)=\Omega^{(0)}J$ at the point of capture, where 
$\Omega^{(0)}=a/(r^2_++a^2)$ is the angular velocity of the black
hole, $J$ is the conserved angular momentum of the particle, 
and $r_+=M$ is the location of the black-hole horizon \cite{Note1}.
             
One should also consider {\it first}-order interactions between the black
hole and the particle's angular momentum. As the particle spirals into the
black hole it interacts with the black hole, so the horizon generators 
start to rotate, such that at the point of assimilation the black-hole
angular velocity, $\Omega$, has changed from $\Omega^{(0)}$ to
$\Omega^{(0)}+\Omega^{(1)}_c$. The corresponding first-order energy
correction is ${\cal E}^{(1)}_{self}=\Omega^{(1)}_cJ$. 
On dimensional analysis one expects $\Omega^{(1)}_c$ to be
of the order of $O(J/M^3)$. In fact, Will \cite{Will} has performed a
perturbation analysis for the problem of a {\it ring} of particles rotating
around a slowly spinning (neutral) black hole, and found
$\Omega^{(1)}_c=J/4M^3$. As would be expected from a perturbative
approach, $\Omega^{(1)}_c$ is proportional to $J$. To our best
knowledge, no exact calculation of $\Omega^{(1)}_c$ has been performed 
for generic (Kerr-Newman) black holes, nor for the case of a {\it single} particle (in 
which case the system loses the axial symmetry which characterized it in the case of a ring of matter). 
We therefore write ${\cal E}^{(1)}_{self}=\omega J^2$, and obtain

\begin{equation}\label{Eq1}
{\cal E}={\cal E}^{(0)}+{\cal E}^{(1)}_{self}={{J} \over {2M}}+ \omega J^2\  ,            
\end{equation}
for the particle's energy at the point of capture.
             
The assimilation of the particle results with a change $\Delta
M={\cal E}$ in the black-hole mass, and a change $\Delta(Ma)=J$ in its
angular momentum. The condition for the black hole to preserve its integrity
after the assimilation of the particle ($a_{new} \leq M_{new}$) is:
 
\begin{equation}\label{Eq2}
{{Ma+J} \over {M+{\cal E}}} \leq M+{\cal E}\  ,
\end{equation}
or equivalently [Substituting ${\cal E}$ from Eq. (\ref{Eq1})]

\begin{equation}\label{Eq3}
0 \leq J^2 \left(4M\omega +{1 \over {2M^2}} \right) \  ,
\end{equation}
which is automatically satisfied. We therefore conclude that the black-hole
horizon cannot be removed by the assimilation of the particle --
cosmic censorship is upheld.

We next consider the case of a particle 
which is lowered towards a {\it near}-extremal Kerr black hole. 
The condition for the black hole to preserve its integrity, Eq. (\ref{Eq2}), yields

\begin{equation}\label{Eq4}
0 \leq \left({J \over M} -\varepsilon \right)^2 + J^2 \left(4M\omega -{1 \over {2M^2}} \right)\  ,
\end{equation}
where $r_{\pm} \equiv M \pm \varepsilon$. Perhaps somewhat surprisingly, the situation is 
more involved than in the extremal case: the test particle
approximation ($\omega =0$) actually allows a {\it near} extremal Kerr black hole 
to ``jump over'' extremality by capturing a particle with angular momentum. One 
must refer to the self-energy of the particle (first-order interactions between the black
hole and the object's angular momentum) in order to insure the validity of the 
cosmic censorship conjecture. In fact, we may reverse the line of reasoning: with the 
plausible assumption of cosmic censorship, it is possible to infer a lower bound on the self-energy 
of the particle: ${\cal E}^{(1)}_{self} \geq J^2/8M^3$.

We next generalize our results to 
the Kerr-Newman case. The 
energy of the particle at the point of capture is now given by 

\begin{equation}\label{Eq5}
{\cal E}={\cal E}^{(0)}+{\cal E}^{(1)}_{self}={{aJ} \over {r^2_++a^2}}+ \omega J^2\  .            
\end{equation}
The black-hole condition $M^2-a^2-Q^2 \geq 0$ (after the assimilation of the particle) now reads

\begin{equation}\label{Eq6}
0 \leq (M+{\cal E})^2 -\left({{Ma+J} \over {M+{\cal E}}}\right)^2 -Q^2\  ,
\end{equation}
which implies

\begin{equation}\label{Eq7}
0 \leq \left({2a \over {M^2+a^2}}J -\varepsilon \right)^2 + J^2
\left({2\omega} {{M^2+a^2} \over M} -{1 \over {M^2+a^2}} \right)\ .
\end{equation}
Thus, one may derive a necessary condition for the validity of the cosmic censorship
conjecture (a lower bound on the self-energy ${\cal E}^{(1)}_{self}$):

\begin{equation}\label{Eq8}
{\cal E}^{(1)}_{self} \geq {M \over {2(M^2+a^2)^2}} J^2 \  .
\end{equation}

Furthermore, if the resulting configuration (after the assimilation of the particle) is 
a black hole, then according to Hawking's {\it area theorem} \cite{Haw} there should be a 
growth (or no change) in the area of the black hole. The surface area, $A$, of a Kerr-Newman black hole is given by 
$A=4\pi(2Mr_+-Q^2)$, where $r_+=M+(M^2-a^2-Q^2)^{1/2}$ is the location of the black-hole outer horizon. 
Substituting $M \to M+{\cal E}$ and $Ma \to Ma+J$, 
one may use the area theorem ($A_{old} \leq A_{new}$) to derive a lower bound on the 
particle self-energy:

\begin{equation}\label{Eq9}
{\cal E}^{(1)}_{self} \geq {r^2_+ \over {2M\alpha^2}} J^2 \  ,
\end{equation}
where $\alpha=A/4\pi$. This bound is valid for any Kerr-Newman black hole (not necessarily a near extremal one).

We note that the bound Eq. (\ref{Eq8}) derived from the cosmic censorship conjecture is stronger 
than the one derived from the area theorem, Eq. (\ref{Eq9}) (There is an equality in the 
extremal limit, where $r_+ \to M$.) Thus, the analysis is self-consistent -- 
provided cosmic censorship is respected, there is a growth in the black-hole surface area.

\section{Self-energy of a charged particle}

We next consider a charged particle of rest mass $\mu$, charge $q$, and proper
radius $R$, which is (slowly) descent into a (near extremal) Kerr black hole. 
The total energy $\cal E$ of the particle in a black-hole spacetime 
is made up of two distinct contributions: $ 1) \  {\cal E}_0$, the energy
associated with the body's mass 
(red-shifted by the gravitational field), and 
$ 2) \ {\cal E}^{(1)}_{self}$, the gravitationally induced self-energy of the charged particle. 

The first contribution, ${\cal E}_0$, is given by Carter's \cite{Carter} integrals 
for a particle moving in a black-hole background:
             
\begin{equation}\label{Eq10}
{\cal E}_0={{\mu \ell (r_+-r_-)} \over {2\alpha}} 
[1+O(\ell^2/{r^2_+})]\  ,
\end{equation}
where $r_{\pm}=M \pm (M^2-a^2)^{1/2}$ are the
locations of the black-hole (event and inner) horizons, and 
$\ell$ is the proper distance from the horizon. Namely,
             
\begin{equation}\label{Eq11}
\ell = \ell(r)= \int_{r_{+}}^{r} \sqrt{g_{rr}} dr\  ,
\end{equation}
with $g_{rr}=(r^2+a^2 \cos^2 \theta)/(r-r_+)(r-r_-)$. 

The second contribution, ${\cal E}^{(1)}_{self}$, reflects the effect of the 
spacetime {\it curvature} on the particle's 
electrostatic {\it self-interaction}. The physical origin of this force 
is the distortion of the charge's long-range Coulomb field by
the spacetime curvature. This can also be interpreted as being due to the
image charge induced inside the (polarized) black hole
\cite{Linet,BekMay}. The self-interaction of a charged particle in the
black-hole background results with a repulsive (i.e., directed away from the
black hole) self-force. A variety of
techniques have been used to demonstrate this effect in black-hole 
spacetimes \cite{DeDe,Ber,Mac,Vil,SmWi,ZelFro,Lohi,LeLi1,LeLi2}. In particular, the contribution of this effect to the
particle's (self) energy in the Schwarzschild background is ${\cal E}^{(1)}_{self}=Mq^2/2r^2$, which implies
${\cal E}^{(1)}_{self}=q^2/8M$ in the vicinity of the black hole. However, in the generic 
case of a spinning Kerr black hole, the 
self-energy was calculated only for the specific case in which the particle is located 
along the symmetry axis of the black hole. 
We therefore write ${\cal E}^{(1)}_{self}=\eta q^2$.

The gradual approach to the black hole must stop when the
proper distance from the body's center of mass to the black-hole
horizon equals $R$, the body's radius. 
One therefore finds

\begin{equation}\label{Eq12}
{\cal E}={{\mu R (r_+-r_-)} \over {2\alpha}}+\eta q^2\  ,
\end{equation}
for the particle's energy at the point of capture (this expression is valid for an 
arbitrary value of the azimuthal angel $\theta$).
 
An assimilation of the charged particle results with a 
change $\Delta M={\cal E}$ in the
black-hole mass, and 
a change $\Delta Q=q$ in its charge. The condition
for the black hole to preserve its integrity after the assimilation of
the charge is therefore

\begin{equation}\label{Eq13}
0 \leq (M+{\cal E})^2 -\left({{M} \over {M+{\cal E}}}\right)^2 -q^2\  ,
\end{equation}
or equivalently

\begin{equation}\label{Eq14}
0 \leq q^2 (4M\eta -1) +2\varepsilon \mu R/M\  .
\end{equation}
We emphasize that Eq. (\ref{Eq14}) implies that the test particle approximation (that is, taking 
$\eta=0$) allows to over-charge a black hole.

The Coulomb energy of a charged particle is given by $fq^2/R$, 
where $f$ is a numerical factor of order unity which depends on how the charge is
distributed inside the body. The Coulomb energy 
attains its minimum, $q^2/2R$, when the charge is uniformly spread on a thin           
shell of radius $R$, which implies $f \geq 1/2$ (an homogeneous
charged sphere, for instance, has $f=3/5$). Therefore, any charged body which respects the   
weak (positive) energy condition must be larger than $r_c \equiv q^2/2\mu$. 
Thus, a necessary condition for the validity of the cosmic censorship conjecture is 
[see Eq. (\ref{Eq14})]

\begin{equation}\label{Eq15}
0 \leq q^2 (4M\eta -1+\varepsilon /M)\  ,
\end{equation}
which implies a lower bound on the self-energy, ${\cal E}^{(1)}_{self}$, of a charged 
particle:

\begin{equation}\label{Eq16}
{\cal E}^{(1)}_{self} \geq {{q^2} \over {4M}} \left(1 -{{\varepsilon} \over {M}} \right)\  .
\end{equation}

We next apply Hawking's area theorem to the gedanken experiment. If the resulting configuration is 
a black hole, then the area theorem \cite{Haw} (namely, $A_{old} \leq A_{new}$) imposes 
a lower bound on the particle self-energy:

\begin{equation}\label{Eq17}
{\cal E}^{(1)}_{self} \geq {M\over {2\alpha}} q^2 \  .
\end{equation}

We note that an exact expression for the self-energy of a charged particle 
is available only for the specific case in which the 
particle is placed along the {\it symmetry} axis ($\theta=0$) of the Kerr black hole \cite{Lohi,LeLi1}: 
${\cal E}^{(1)}_{self}={Mq^2 /{2\alpha}}$. Note that this result coincides exactly with the bound 
Eq. (\ref{Eq17}). Furthermore, the exact result (available only in the $\theta=0$ case) yields 
${\cal E}^{(1)}_{self}={q^2 \over {4M}}(1 -\varepsilon /M)$ for 
a near extremal Kerr black hole. Thus, taking cognizance of Eq. (\ref{Eq16}) we find 
that cosmic censorship is respected provided one 
takes into account the electrostatic self-energy of the particle in the background of the black hole.

In summary, in this paper we have analyzed gedanken experiments in which particles 
carrying angular momentum or electric charge are assimilated by a black hole. 
The gedanken experiments are considered from the point of view of Penrose 
cosmic censorship conjecture and Hawking's area theorem. 
It was shown that first-order interaction effects (the self-energy of the particle) 
{\it must} be taken into account in order to preserve the black-hole integrity and to 
insure the validity of the cosmic censorship conjecture. 

Moreover, exact calculations of the self-energy are available in 
the literature only for a limited number of cases. Using Hawking's area theorem, 
we derived bounds on the self-energy of a 
particle in the vicinity of a black hole. The resulting bounds are summarized in 
Table \ref{Tab1}.

\bigskip
\noindent
{\bf ACKNOWLEDGMENTS}
\bigskip

I would like to thank Mordehai Milgrom and Jacob D. Bekenstein for helpful discussions. 
This research was supported by grant 159/99-3 from the Israel 
Science Foundation and by the Dr. Robert G. Picard Postdoctoral Fellowship in 
Physics.

\begin{table}
\caption{Self-energy of a particle in the vicinity of a black hole.}

\label{Tab1}
\begin{tabular}{llcc}
Type of self-energy &Type of a black hole & Lower bound on self-energy & Exact calculation\\
\tableline
Rotational & Kerr-Newman & ${r^2_+ \over {2M\alpha^2}} J^2$ & ? \\
Electrostatic & Kerr (symmetry axis)& ${M\over {2\alpha}} q^2$ & ${M \over {2\alpha}} q^2$ \\
Electrostatic & Kerr ($\theta \neq 0$)& ${M\over {2\alpha}} q^2$ & ? \\
\end{tabular}
\end{table}

\end{document}